\documentclass[aps,preprint,amsmath,amssymb]{revtex4}
\usepackage{graphicx,color}
\usepackage{subfigure}
\def\be{\begin{eqnarray}}
\def\ed{\end{eqnarray}}
\def\non{\nonumber}

\def\lam{\lambda}

\begin{document}


\title{Search for $\delta^{\pm\pm}$ with new decay patterns at the LHC}

\author{ \bf Chuan-Hung Chen$^{a}$\footnote{Email:
physchen@mail.ncku.edu.tw} and Takaaki Nomura$^{a}$\footnote{Email: nomura@mail.ncku.edu.tw} }

\affiliation{ $^{a}$Department of Physics, National Cheng-Kung
University, Tainan 701, Taiwan  }

\date{\today}

\begin{abstract}

 A study of searching for doubly charged Higgs $(\delta^{\pm\pm})$ is performed in two-Higgs-doublet extension of the conventional type-II seesaw model. We find that a fantastic mixing effect between singly charged Higgs of Higgs doublet and of triplet is arisen from the scalar potential. The mixing leads to  following  intriguing phenomena:  (a) the mass splittings in triplet particles are magnified, (b) QCD processes dominate the production of $\delta^{\pm\pm}$, and (c) new predominant decay channels of $\delta^{\pm\pm}$ are $\delta^{\pm\pm} \to W^{\pm^{[*]}} H^{\pm^{(*)}}_{1(2)}$, but not $\delta^{\pm}\to (\ell^\pm \ell^\pm, W^\pm W^\pm)$ which are usually discussed in the literature. With luminosity of 40 fb$^{-1}$ and collision energy of 13 TeV, we demonstrate that $\delta^{\pm\pm}$ with mass below $330$ GeV could be observed at the $5\sigma$ level. Moreover,  when the luminosity approaches to 300 fb$^{-1}$, the observed  mass of $\delta^{\pm\pm}$ could reach up to 450 GeV. 

\end{abstract}

\maketitle

\section{Introduction}

The origin of the masses of standard model (SM) particles  from spontaneous electroweak symmetry breaking,  called Higgs mechanism, is supported  by the new observed scalar boson with a mass around 125 GeV, measured by ATLAS~\cite{:2012gk} and CMS~\cite{:2012gu}. 
Following the same concept,   the mystery of  tiny neutrino masses may be solved in the framework of multiple  Higgs fields  without introducing the heavy singlet right-handed neutrinos \cite{SeeSaw}, in which  the representative model is the Higgs triplet extension of the SM \cite{Magg:1980ut} and here we call it as the conventional type-II seesaw model (CTTSM). 

The novel feature of a Higgs triplet model is the existence of a doubly charged Higgs, hereafter denoted by $\delta^{\pm\pm}$. Therefore, in order to detect the Higgs triplet particles, the searches for $\delta^{\pm\pm}$ at colliders have been studied widely by theorists~\cite{Han:2007bk, Akeroyd:2007zv,delAguila:2008cj,Aoki:2011pz,Chiang:2012dk,Sugiyama:2012yw,Kanemura:2013vxa,Chun:2013vma,Akeroyd:2005gt,Melfo:2011nx,Chun:2003ej,Perez:2008ha,Arhrib:2011uy,Dutta:2014dba} and experimentalists~\cite{Chatrchyan:2012ya,ATLAS:2012hi,Khachatryan:2014sta,Dev:2013ff}. The collider signatures ordinarily depend on the vacuum expectation value (VEV) of the neutral triplet field, $v_\Delta$, which  is the source of neutrino masses. For instance, if $v_\Delta \ll 10^{-4}$ GeV or the associated leptonic Yukawa couplings are relatively large,  it is found that the doubly charged Higgs mainly decays into a pair of same-sign charged leptons, i.e. $\delta^{\pm\pm} \to \ell^{\pm} \ell^\pm$ ($\ell=e,\, \mu$)~\cite{Han:2007bk, Perez:2008ha}. However, if $v_\Delta \gg 10^{-4}$ GeV or the associated leptonic Yukawa couplings are relatively small, the dominant decay channel of $\delta^{\pm\pm}$ is $\delta^{\pm\pm} \to W^\pm W^\pm$~\cite{Han:2007bk, Perez:2008ha}. Therefore, the searches for $\delta^{\pm \pm}$ at colliders usually are focused on the decay channels of same-sign leptons or same-sign W bosons. Consequently, if one assumes that $\delta^{\pm \pm}$ is 100\% decaying into leptons, the experimental lower bound of its mass now is around 400 GeV~\cite{Chatrchyan:2012ya, ATLAS:2012hi}. If $WW$ channel is dominant, the mass limit of $\delta^{\pm\pm}$ now is up to  300 (550) GeV when $v_\Delta = 25 (35)$ GeV ~\cite{Khachatryan:2014sta}.

 Historically, the two-Higgs-doublet model (THDM) was proposed  for solving the weak and strong CP problems~\cite{Lee:1973iz,Peccei:1977hh}. In spite of the original motivation, THDM itself provides rich phenomena in particle physics. By the new discovery of  125 GeV scalar boson at ATLAS and CMS, the phenomenology of THDM has been further investigated broadly in the literature, e.g. Refs.~\cite{ Ferreira:2012nv,Diaz-Cruz:2014aga,Barger:2014qva}.  By combining the issue of neutrino physics, it is intriguing to explore the fantastic effects in the model involving two Higgs doublets (THDs) and one Higgs triplet. Indeed, we find that the production and decay patterns of doubly charged Higgs will be completely changed when the second Higgs doublet is added to the CTTSM. 

In the THD type-II seesaw model, although we have new interacting terms from various sectors, the most attractive new effects are  the dimension-3 terms in scalar potential, read by $\mu_j H^T_j i\tau_2 \Delta^\dagger H_k$ (j,k =1,2), where $H$ and $\Delta$ are the Higgs doublet and triplet, respectively. Since the coefficients $\mu_j$ are of order of electroweak (EW) scale, the new terms lead a large mixing angle  between the singly charged Higgs of doublet ($H^\pm$) and of triplet ($\delta^{\pm}$). According to  our previous study~\cite{Chen:2014xva},  if we assume the Higgs triplet particles are heavier than the Higgs doublets, due to the new mixing effect,  we have the interesting phenomena: (I) the charged Higgs $H^\pm$ could be lighter than that in type-II THDM, (II) even we set the degeneracy of $m_{\delta^{\pm\pm}} = m_{\delta^\pm}$, the mass splitting between $\delta^{\pm\pm}$ and $\delta^{\pm}$ could be magnified, (III) the branching fractions for $\delta^{\pm\pm} \to W^\pm H^\pm_i$ are much larger than those for $\delta^{\pm\pm} \to ( \ell^\pm \ell^\pm, W^\pm W^\pm)$. Due to these new characters, one expects that the signals of $\delta^{\pm\pm}$ in  the THD type-II seesaw model are different from those signals in other triplet  models. 

For exploring the signals of $\delta^{\pm\pm}$, in this paper we study its various production processes. 
Since now the Higgs doublets could couple to the Higgs triplet, unlike the cases in CTTSM and Georgi-Machacek model \cite{Georgi:1985nv,Chiang:2012dk} where EW processes dominate, we find that the doubly charged Higgs in our model is predominantly 
produced by QCD processes, indicated by $pp \to H_2^+ \bar t b (H_2^- t \bar b) \to \delta^{++} W^- \bar t b (\delta^{--} W^+ t \bar b)$ and $\bar b(b) g \to H_2^+ \bar t (H_2^- t) \to \delta^{++}W^- \bar t (\delta^{--} W^+ t)$.  Due to more jets involved in the production and decays of $\delta^{\pm\pm}$, the selected events for simulation are $\ell^{\pm} \ell^{\pm}$ + $n$jets with $n\geq 4$. For reducing the possible background events, we propose several kinematical cuts on the second highest transverse-momentum lepton and the invariant mass of the same-sign dilepton.  Additionally, we also study the discovery potential for $5\sigma$ significance   with the collision energy of 13 TeV and the designed luminosity at the LHC.

We organize the paper as follows. In Sec.~II, we briefly discuss the relevant new interactions originated in Yukawa sector, gauge invariant kinetic terms of involved scalar fields and the scalar potential. The new characters of doubly charged Higgs is also introduced. We investigate the production and decays of doubly charged Higgs and the branching fractions of singly charged Higgses in Sec.~III. The detailed simulation on signals and backgrounds are given in Sec.~IV. We summarize the findings in Sec.~V. 

\section{ New characters of doubly charged Higgs }

For studying the  detection of doubly charged Higgs  in the THD  and one Higgs triplet model, we  first summarize the relevant interactions with $\delta^{\pm\pm}$. The detailed introduction to the model could refer to Ref.~\cite{Chen:2014xva}. For satisfying the gauge symmetry of the SM, $\delta^{\pm\pm}$  can  only directly couple to leptons  in leptonic Yukawa sector and the couplings are expressed by
\be
 {\cal L}_{\delta^{\pm\pm} \ell \ell} &=& \frac{1}{2} \ell^{\prime T} C {\bf h} P_L \ell' \delta^{++} +h.c. \,, \non \\
 {\bf h}&=& =\frac{\sqrt{2}}{v_{\Delta}} U^*_{\rm PMNS} {\bf m}^{\rm dia}_{\nu}U^{\dagger}_{\rm PMNS} \,,
 \label{eq:int_y}
 \ed
where $\ell^{\prime}$ denotes the charged leptons,  ${\bf m}^{\rm dia}_\nu$ is the diagonalized neutrino mass matrix and $U_{\rm PMNS}$ is the Pontecorvo-Maki-Nakagawa-Sakata (PMNS) matrix \cite{Pontecorvo:1957cp,Maki:1962mu}. From Eq.~(\ref{eq:int_y}), one can see that the typical coupling of $\delta^{\pm\pm}$ to lepton-pair is proportional to $m_\nu /v_\Delta$. If we assume that the masses of neutrinos are measured well in experiments, the partial decay rates for $\delta^{\pm\pm}\to \ell^{\prime\pm} \ell^{\prime\pm}$  depend on the value of $1/v_\Delta$. By the gauge invariant kinetic terms of Higgs triplet, the couplings of $\delta^{\pm\pm}$ to gauge bosons are written as 
\be
{\cal L}_{\delta^{\pm\pm} GG } &=& -i \frac{g}{\cos\theta_W} (1- 2 \sin^2\theta_W)  \delta^{--} (\partial_\mu \delta^{++})   Z^\mu - i 2 e \delta^{--} (\partial_\mu \delta^{++}) A^\mu \non \\
&& -ig (\partial_\mu \delta^{++}) \delta^{-} W^{-\mu} + i g \delta^{++} (\partial_\mu \delta^-) W^{-\mu} + \frac{g^2 v_\Delta}{\sqrt{2}} \delta^{++} W^-_\mu W^{-\mu} + h.c.\,, \label{eq:d++W}
\ed
where $\delta^{\pm}$ are the singly charged Higgs of Higgs triplet.  Clearly, the branching fraction for $\delta^{\pm\pm} \to W^\pm W^\pm$ depends on the magnitude of $v_\Delta$. According to Eq.~(\ref{eq:int_y}) and Eq.~(\ref{eq:d++W}), one can realize that in CTTSM, the main decays of $\delta^{\pm\pm}$ are through the two-body decays $\delta^{\pm\pm} \to (\ell ^\pm \ell^\pm, W^\pm W^\pm)$ and three-body decays $\delta^{\pm\pm} \to W^{\pm^{(*)}} \delta^{\pm^{(*)}} $, in which the off-shell condition relies on the mass of $\delta^\pm$.

Since there is only one $\delta^{++(--)}$ in the model, the property  changes  of $\delta^{\pm\pm}$ are arisen from the new interacting terms in    scalar potential. In order to clearly understand the effects, we write the  gauge invariant scalar potential as 
 \be
 V(H_1, H_2, \Delta) &=&  V_{H_1H_2} + V_\Delta + V_{H_1 H_2 \Delta}\,, \non \\
V_{H_1H_2} &=& m^2_1 H^\dagger_1 H_1 + m^2_2 H^\dagger_2 H_2 - m^2_{12} ( H^\dagger_1 H_2 + h.c.)+ \lam_1 ( H^\dagger_1 H_1)^2  \non \\
 &+&  \lam_2 (H^\dagger_2 H_2)^2 + \lam_3  H^\dagger_1 H_1 H^\dagger_2 H_2+ \lam_4  H^\dagger_1 H_2 H^\dagger_2 H_1 + \frac{\lam_5}{2} \left[(H^\dagger_1 H_2)^2+h.c. \right] \,, \non \\
 V_\Delta &=& m^2_\Delta Tr \Delta^\dagger \Delta + \lam_{9} (Tr \Delta^\dagger \Delta)^2 + \lam_{10} Tr (\Delta^\dagger \Delta)^2\,, \non \\
 V_{H_1H_2\Delta} &=& \left( \mu_1 H^T_1 i\tau_2 \Delta^{\dagger}  H_1 + \mu_2 H^T_2 i \tau_2 \Delta^\dagger H_2 + \mu_3 H^T_1 i\tau_2 \Delta^\dagger H_2 + h.c. \right) \non \\
 &+& \left( \lam_6 H^\dagger_1 H_1 + \bar\lam_6 H^\dagger_2 H_2 \right) Tr \Delta^\dagger \Delta + H^\dagger_1 \left( \lam_7 \Delta \Delta^\dagger + \lam_8 \Delta^\dagger \Delta \right) H_1 \non \\
 &+& H^\dagger_2 \left( \bar\lam_7 \Delta \Delta^\dagger + \bar\lam_8 \Delta^\dagger \Delta \right) H_2\,, \label{eq:v}
  \ed
 where $V_{H_1 H_2}$ and  $V_\Delta$  stand for the scalar potential of THD and of pure triplet, and  $V_{H_1 H_2 \Delta}$ is the part involving  $H_1$, $H_2$ and $\Delta$.  By taking the VEVs of $H_{1,2}$ and $\Delta$ to be $v_{1,2}$ and $v_\Delta$ respectively, the vacuum stability requires  
  \be
  v_\Delta \approx \frac{1}{\sqrt{2}} \frac{\mu_1 v^2_1 + \mu_2 v^2_2 + \mu_3 v_1 v_2}{m^2_\Delta + (\lam_6+\lam_7) v^2_1/2 + (\bar\lam_6+\bar\lam_7)v^2_2 /2}\,. \label{eq:v_d}
  \ed
Due to the precision measurement  of $\rho$-parameter, we have  $v_\Delta \ll  v=\sqrt{v^2_1+ v^2_2}$ and just keep the leading power for $v_\Delta$ in Eq.~(\ref{eq:v_d}). By this result, we see that  when $\mu_2=\mu_3 =0$, the small $v_\Delta$ indicates  the small $\mu_1$ or large  $m_\Delta$ in CTTSM.  However, when the $\mu_2$ and $\mu_3$ effects are introduced, the necessity  of small $v_\Delta$ could be accommodated by  the massive parameters $\mu_{1,2,3}$ and $m_\Delta$, where they could be in the same order of magnitude. Hence, the small value of  $v_\Delta$  could be adjusted by the free parameters of the new scalar potential,  
without introducing a hierarchy to the massive parameters. 

It is known that in CTTSM, the mixing effect of Higgs doublet and triplet is related to the suppressed factor  $v_\Delta / v$. However, an interesting mixing effect could be induced in the THD extended type-II seesaw model when $\mu_{1,2,3}$ in Eq.~(\ref{eq:v}) are all as large as  EW scale.  For displaying the influence of $\mu_{1,2,3}$,  we take the singly charged Higgses as the illustrator. The similar discussions are also suitable for neutral scalars \cite{Chen:2014xva}. As known,  one physical charged Higgs $H^\pm$ exists in the conventional THD model and   a massive Higgs triplet  provides a singly charged Higgs $\delta^\pm$. If we take the approximation of $v_\Delta/v \approx 0$,  we find that the mixture of $\delta^\pm$ with charged Goldstone boson of THD could be ignored.  For simplifying the analysis and preserving the requirement of $v_\Delta << v$, in the numerical estimates, we adopt the relation 
 \be
 \mu_3 \sim - \frac{\mu_1 v^2_1 + \mu_2 v^2_2  }{v_1 v_2}\,. \non 
 \ed
 
By Eq.~(\ref{eq:v}) and decoupling from the Goldstone boson,  the mass matrix of singly charged Higgses in our model  could be formulated  by a $2\times 2$ matrix and expressed by 
\begin{equation}
  ( H^- \delta^- ) \left( \begin{array}{cc}
      m^2_{H^{-}H^{+}} & m^2_{H^- \delta^+} \\ 
 m^2_{H^- \delta^+} & m^2_{\delta^- \delta^+} \\ 
  \end{array} \right)  \left(\begin{array}{c}  
    H^+  \\ 
    \delta^+ \\ 
  \end{array} \right)\,, \label{eq:vm2}
 \end{equation}
where the elements of mass matrix are found by
 \begin{eqnarray}
 m^2_{H^- H^+} &\equiv& m^2_{H^\pm} = \frac{m^2_{\pm}}{\sin\beta \cos\beta}\,,~m^2_{\pm} = m^2_{12}-\frac{\lambda_4 + \lambda_5}{2}v_1 v_2\,,\nonumber \\
m^2_{H^- \delta^+} &=& \frac{v}{2\sin\beta \cos\beta} \left[\mu_1 \cos^4\beta -\mu_2 \sin^4\beta + (\mu_1 -\mu_2 ) \sin^2\beta \cos^2\beta \right] \,, \nonumber \\
m^2_{\delta^- \delta^+} &\equiv& m^2_{\delta^\pm} = m^2_\Delta +\frac{v^2_1}{4} (2\lambda_6 + \lambda_7 + \lambda_8) + \frac{v^2_2}{4} (2\bar\lambda_6 + \bar\lambda_7 + \bar\lambda_8)\,.
 \end{eqnarray}
We see that the off-diagonal element is associated with the parameters $\mu_{1,2}$ and $\tan\beta = v_2/v_1$. 
The physical charged Higgs states could be regarded as the combination of $H^\pm$ and $\delta^\pm$ and their mixture  could be parametrized by 
 \be
  \left( \begin{array}{c}
    H^\pm_1\\ 
    H^\pm_2 \\ 
  \end{array}\right) =   \left(\begin{array}{cc}
    \cos\theta_\pm & \sin\theta_\pm \\ 
    -\sin\theta_\pm & \cos\theta_\pm\\ 
  \end{array}\right) \left( \begin{array}{c}
    H^\pm \\ 
    \delta^{\pm} \\ 
  \end{array}\right)\,.\label{eq:ma}
 \ed 
The masses of charged Higgs particles and their mixing angle are derived as
  \be
 \left( m_{H_{1,2}^{\pm}}\right)^2 &=&  \frac{1}{2}\left(m^2_{\delta^\pm} + m^2_{H^\pm}\right) \mp \frac{1}{2} \left[ \left( m^2_{\delta^\pm} - m^2_{H^\pm}\right)^2 + 4 m^4_{H^-\delta^+}\right]^{1/2}\,, \non \\
 \tan2\theta_\pm &=& - \frac{2 m^2_{H^- \delta^+}}{m^2_{\delta^\pm} - m^2_{H^\pm}}\,.
 \label{eq:mass_mixing}
  \ed
Here $H^\pm_1$  is identified as the lighter charged Higgs. Clearly, the magnitude of mixing angle $\theta_\pm$ relies on the massive parameters $\mu_{1,2}$. In this paper, we are going to explore the influence of large mixing angle $\theta_\pm$ on the search for the doubly charged Higgs. With the new mixing effect,   we present the couplings of  $\delta^{\pm \pm}$ to the physical states $H^{\pm}_{i}$ and $W^{\pm}$  in Table~\ref{tab:interactions}. By the Table, we see that the involved free parameter for the vertex  $\delta^{\pm\pm}$-$H^\mp_{i}$-$W^\mp$ is only  the angle $\theta_\pm$.  Although the coupling for the vertex $\delta^{\pm\pm}$-$H^\mp_{i}$-$H^\mp_{i}$ could be comparable with that for  $\delta^{\pm\pm}$-$H^\mp_{i}$-$W^\mp$, due to phase space suppression,  the decay rate for $H^\pm_{i} H^\pm_i$ mode usually will be smaller than that for $H^\pm_i W^\pm$ mode, except the case of $\tan\beta=1$ with $\mu_1=\mu_2$ and the case constrained by kinematic requirement~\cite{Chen:2014xva}.  
%
\begin{table}[hpbt]
\begin{ruledtabular}
\begin{tabular}{cc|cc}
Vertex & Coupling & Vertex & Coupling \\ \hline \hline
 $\delta^{\pm \pm} H_2^\mp W_\mu^\mp $ & $-ig \cos \theta_\pm (p_{\delta^{\pm \pm}} - p_{H_2^\mp})_\mu$ 
 & $\delta^{\pm \pm} H_1^\mp W_\mu^\mp $ & $-ig \sin \theta_\pm (p_{\delta^{\pm \pm}} - p_{H_1^\mp})_\mu$ \\ \hline
 $\delta^{\pm \pm} H_{1(2)}^{\mp} H_{1(2)}^{\mp}$ & $2 (\mu_1+\mu_2) \cos^2 \theta_\pm (\sin^2 \theta_\pm)$ 
 & $\delta^{\pm \pm} H_1^{\mp} H_2^{\mp}$ & $2 (\mu_1+\mu_2) \cos \theta_+ \sin \theta_+$ 
\end{tabular}
\caption{The couplings of $\delta^{\pm\pm}$ to $H^{\pm}_{1,2}$ and $W^\pm$. 
 \label{tab:interactions}}
 \end{ruledtabular}
\end{table}
%

Although $\delta^\pm$ cannot couple to quarks directly, however due to the new mixing effect in Eq.~(\ref{eq:ma}), the two physical charged Higgses now can interact with quarks and the interactions with fermions are formulated by
 \be
 {\cal L}_{H^\pm_i  f f'} &=& \frac{\sqrt{2}}{v}  \left[\bar u\left( \tan\beta V_{CKM} {\bf m_D}   P_R + \cot\beta  {\bf m_U} V^\dagger_{CKM} P_L \right) d + \tan\beta \bar\nu {\bf m_\ell} P_R \ell' \right] \non \\
 &\times& (\cos\theta_\pm H^+_1 -  \sin\theta_\pm H^+_2) + h.c. \,, \label{eq:hff}
  \ed
 where we suppress all flavor indices, $u^T=( u, c, t)$ and $d^T = ( d, s, b)$ denote the up and down type quarks, $\nu^T= (\nu_e, \nu_\mu, \nu_\tau)$ and $\ell^{\prime T}=( e, \mu ,\tau)$ are the neutrinos and charged leptons, $V_{CKM}$ is the Cabibbo-Kobayashi-Maskawa (CKM) matrix, ${\bf m_{D(U)}}$ is the diagonalized mass matrix of down (up) type quarks, and $P_{R, L} = (1 \pm \gamma_5)/2$. Note that the Yukawa couplings of $\delta^+$ and leptons are assumed to be small and negligible, thus we do not show them in  Eq.~(\ref{eq:hff}). 
  
 \section{ Production and decays of doubly charged Higgs}
 
 In order to search for the signals of doubly charged Higgs, we need to understand its producing mechanisms and the main decay modes.  In the following discussions, we focus on the production of $\delta^{\pm\pm}$ and its  decays.

 \subsection{ Production of doubly charged Higgs at LHC}
 
 %
 According to the interactions in Eq.~(\ref{eq:d++W}) and Table~{\ref{tab:interactions}},  we see that the $\delta^{\pm\pm}$ could be produced 
by EW interactions via s-channel, read as
 \begin{align}
& p p \rightarrow Z/\gamma \rightarrow \delta^{++} \delta^{--} \,, \label{eq:EW1}\\
& p p \rightarrow W^\pm \rightarrow \delta^{\pm \pm} H_{1,2}^{\mp} \,. \label{eq:EW2}
\end{align}
Except the new mixing effect $\theta_\pm$, the production channels are similar to  those in CTTSM. We note that due to $v_\Delta \ll v$ in our model, the $WW$ fusion is small and negligible. Moreover, with the new effects arisen from $\mu_i$ terms in scalar potential, the on-shell $\delta^{\pm\pm}$ could be produced through the QCD interactions and the relevant  processes are  given by
 \begin{align}
 \label{eq:QCDI}
 & p p \rightarrow H_2^+ \bar{t} b (H_2^- t \bar b) \rightarrow \delta^{++} W^- \bar{t} b (\delta^{--} W^+ t \bar b) \,,  \\
 \label{eq:QCDII}
& p p \rightarrow H_2^+ \bar{t}(H_2^- t)  \rightarrow \delta^{++} W^- \bar{t} (\delta^{--} W^+ t)\,. 
 \end{align}
 Since  the adopted mass relation is $m_{H^\pm_1} < m_{\delta^{\pm \pm}} < m_{H^\pm_2}$, the on-shell doubly charged Higgs in  Eq.~(\ref{eq:QCDI}) and (\ref{eq:QCDII})  is generated by the decay $H^\pm_2 \to \delta^{\pm\pm} W^{\mp}$, and  then follows the decay $\delta^{\pm\pm} \to H_1^\pm W^\pm$.
 The production of $\delta^{\pm\pm}$ through lighter charged Higgs $H^\pm_1$ is  off-shell  effects and small, we therefore ignore its contributions. For the processes in Eq.~(\ref{eq:QCDII}), the main QCD reaction is associated with the interactions of b-quark and gluons, e.g. $\bar b(b) g \to H^+_2 \bar t ( H^-_2 t)$.

Besides the mass parameters of $H^\pm_{1,2}$ and $\delta^{\pm\pm}$, the involved new free parameters for  $\delta^{\pm\pm}$ production are $\sin\theta_\pm$ and $\tan\beta$. It is known that  $\delta^{\pm}$ and $\delta^{\pm\pm}$ belong to the same triplet state, after electroweak symmetry breaking, as expected that their mass splitting should be of order of EW scale. For reducing the number of free parameters and guaranteeing to have a positive definite $m_{H^\pm_{1,2}}$ shown in Eq.~(\ref{eq:mass_mixing}), instead of scanning over the parameter spaces, we set the correlations of parameters for numerical analysis  to be 
\begin{align}
& m_{\delta^\pm} = m_{\delta^{\pm \pm}} + 100 \ {\rm GeV} \,, \non \\
& m_{H^\pm} = \frac{4}{5} m_{\delta^{\pm}} \,, \non \\
& \mu_1 = -\mu_2 = m_{\delta^\pm} \sin \beta \cos \beta \,, \label{eq:setting}
\end{align}
where the setting of $\mu_1 = -\mu_2$ leads  the couplings of $\delta^{\pm\pm} H^\mp_i H^\mp_j$ in Table~\ref{tab:interactions}  to vanish. Accordingly, the masses of charged Higgs and their mixing angle are obtained by
 \begin{align}
& (m_{H_{1,2}^\pm})^2 =  m_{\delta^\pm}^2 \left( \frac{41}{50} \mp \frac{1}{2} \left[ \frac{81}{625} + \frac{v^2}{m_{\delta^\pm}^2} \right]^\frac{1}{2} \right) \,, \non \\
& \tan 2 \theta_\pm = - \frac{25 v}{9 m_{\delta^\pm}}\,. \label{eq:set}
 \end{align}
By the simplified formulae, one can see that the new free parameters are reduced to be $m_{\delta^\pm}$ and $\tan\beta$. 
With the parameter settings of  Eq.~(\ref{eq:setting}),  we find that not only the mass relation $m_{H^\pm_1} < m_{\delta^{\pm \pm}} < m_{H^\pm_2}$ can be satisfied, but also the mixing angle $\theta_\pm$ in Eq.~(\ref{eq:set})  can be large if  $m_{\delta^\pm}$ is of  $O(100)$ GeV. 

\begin{figure}[t] 
\begin{center}
\includegraphics[width=4.0 in]{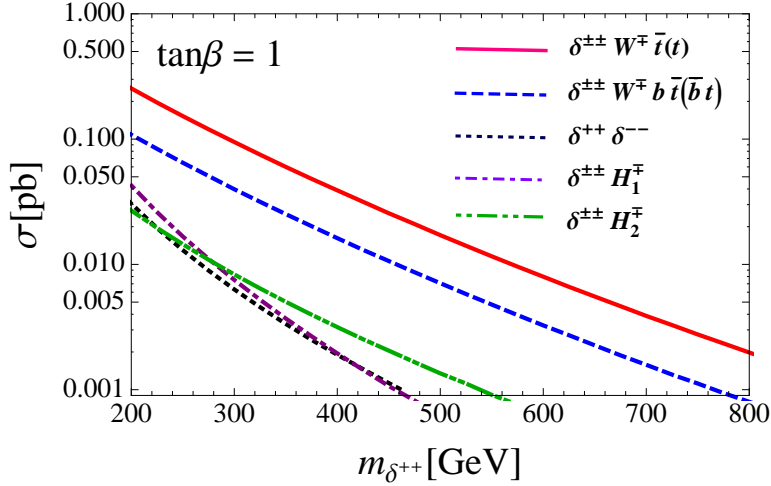} 
\end{center}
\caption{ Production  cross sections of the doubly charged Higgs as a function of $m_{\delta^{\pm \pm}}$ for collision energy of 13 TeV at the LHC, where the  dotted, dash-dotted and dash-dot-dotted lines denote the EW processes while the solid and dashed lines stand for QCD processes.  The settings in Eq.~(\ref{eq:setting}) and $\tan \beta=1$ are adopted. 
\label{fig:XS}}
\end{figure}

\begin{figure}[t] 
\begin{center}
\includegraphics[width=4.0 in]{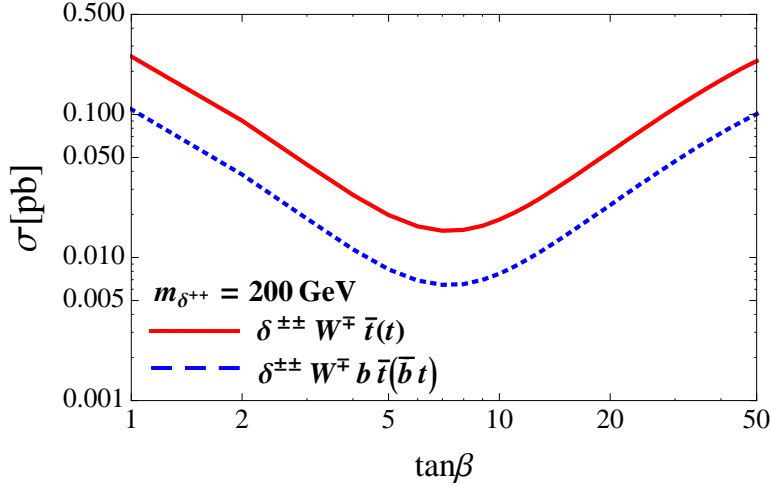} 
\end{center}
\caption{  Production cross sections of $\delta^{\pm\pm}$ as a function of $\tan \beta$ for collision energy of 13 TeV at the LHC, 
where the solid and dashed lines are the QCD processes in Eq.~(\ref{eq:QCDI}) and (\ref{eq:QCDII}), respectively. The settings of Eq.~(\ref{eq:setting}) and $m_{\delta^{\pm\pm}}=200$ GeV are used.  
\label{fig:XStanB}}
\end{figure}

For calculating the production cross section of $\delta^{\pm\pm}$, we employ the {\tt CalcHEP\,3.6.15} code~\cite{CalcHEP} by implementing the parameters and vertices of our model. With the settings of Eq.~(\ref{eq:setting}), $\tan\beta =1$ and the results of Eq.~(\ref{eq:set}), we present the production cross sections for the processes  in Eqs.~(\ref{eq:EW1})-(\ref{eq:QCDII})  as a function of $m_{\delta^{\pm\pm}}$ in  
Fig.~\ref{fig:XS},  where  the collision energy at LHC is 13 TeV and {\tt CTEQ6L} PDF~\cite{Nadolsky:2008zw} is applied, the dotted, dash-dotted and dash-dot-dotted lines denote the EW processes while the solid and dashed lines stand for QCD processes, respectively. 
%
%
Since the contributions of gluons are dominant at $pp$ collision, as expected, the results of QCD production processes are much larger than those of EW ones. 
For further displaying the influence of $\tan\beta$, we fix $m_{\delta^{\pm\pm}} = 200$ GeV and  plot the production cross sections of $\delta^{\pm\pm}$ for QCD processes as a function of $\tan\beta$ in Fig.~\ref{fig:XStanB}.
By the results,  we find that the cross section has a minimum and  occurs at around $\tan\beta=7$.  The larger production cross sections  occur at $\tan\beta \sim O(1)$ or $\sim O(m_t/m_b)$. Based on this result, we concentrate on $\tan\beta =1$ in our numerical calculations. 
%
 
\subsection{ Branching fractions of $\delta^{\pm\pm}$ and charged Higgs}
 
\begin{figure}[t] 
\begin{center}
\includegraphics[width=4.0 in]{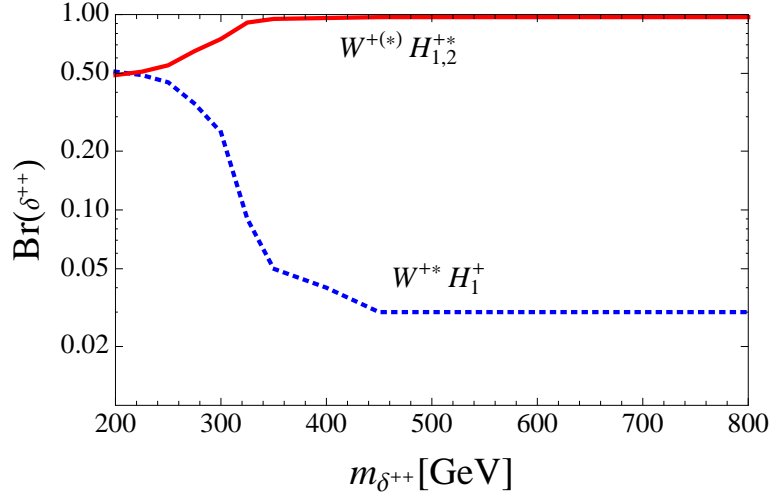} 
\end{center}
\caption{  Branching ratios of doubly charged Higgs decays as a function of $m_{\delta^{\pm\pm}}$. 
\label{fig:Brdcc}}
\end{figure}

When the information for  $\delta^{\pm\pm}$ production is obtained, we then discuss how the doubly charged Higgs decays. According to the introduced interactions, we know that $\delta^{\pm\pm}$ could decay into $\ell^\pm \ell^\pm$, $W^\pm W^\pm$, $H^\pm_i H^\pm_j$, $W^\pm H^\pm_j$, etc. For fitting the tiny masses of neutrinos, if we adopt  the Yukawa couplings of leptons and triplet, $Y_{\ell \ell'}$, and $v_\Delta$  to be small simultaneously, then the first two channels could be ignored. Unlike other Higgs triplet models which only focus on either large $Y_{\ell \ell'}$ or large $v_\Delta$, the suppression of lepton-pair and $W$-pair decays is the new character of our model. With $\mu_1 = -\mu_2$ scheme, the third channel vanishes. Hence, $\delta^{\pm\pm}$ mainly decays into $W^{\pm^{(*)}} H^{\pm^{(*)}}_i$, where $W^\pm$ and $H^{\pm}_i$ could be on-shell and off-shell, depending on the mass of $\delta^{\pm\pm}$.  With the parameter settings of Eq.~(\ref{eq:setting}), we present the branching ratios (BRs) for $\delta^{++} \to W^{+^{(*)}}  H^{+^{(*)}}_{1,2}$ in Fig.~\ref{fig:Brdcc}, 
where the solid line denotes  the BRs for the three (four) body decays of $\delta^{++} \to W^{+^{(*)}}  H^{+^{*}}_{1,2}$  and the dashed line is the BR for the three-body decay of $\delta^{++} \to W^{+^{*}}  H^{+}_{1}$. Due to our parameter settings, the decays for both on-shell $W^\pm$ and $H^\pm_1$ are suppressed. 
By the figure, we see that  when  $m_{\delta^{++}} \gtrsim 300$ GeV, the decays with on-shell $W$-boson become dominant. 

\begin{figure}[hpbt] 
\begin{center}
\includegraphics[width=3.0 in]{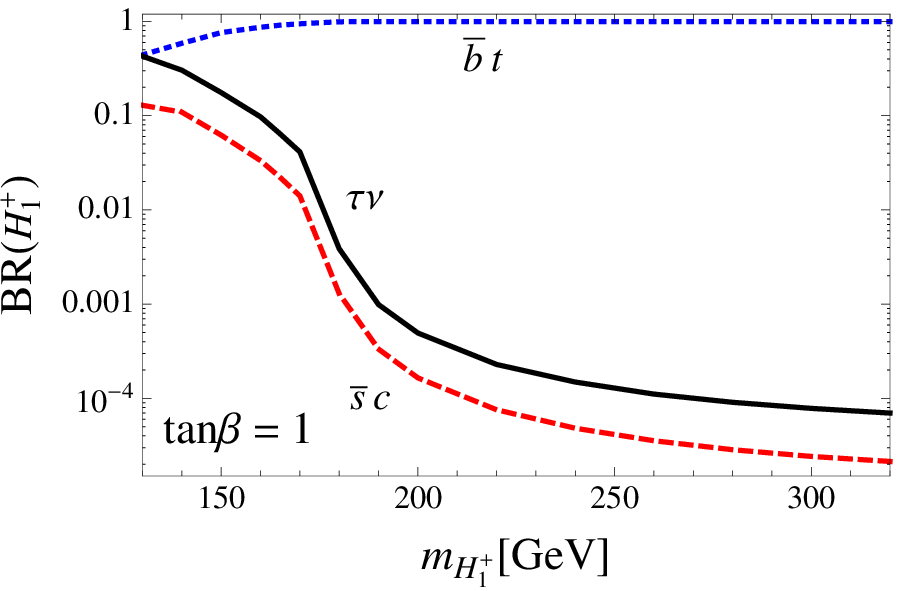} ~\includegraphics[width=3.0 in]{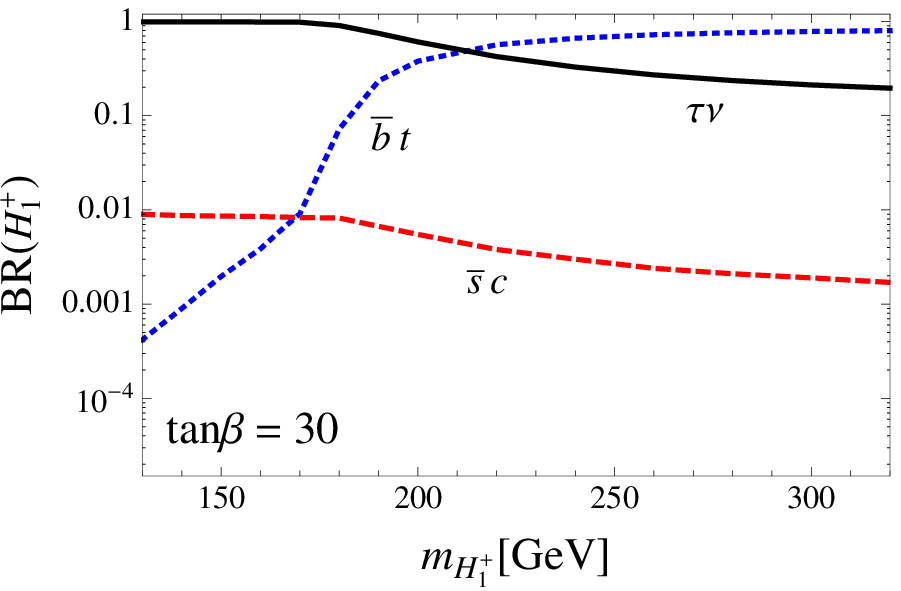}
\end{center}
\caption{  Branching ratios for lighter charged Higgs decays as a function of $m_{H^\pm_1}$ with $\tan\beta=1$ (left panel) and $\tan\beta=30$ (right panel). 
\label{fig:BrHc}}
\end{figure}

\begin{figure}[hpbt] 
\begin{center}
\includegraphics[width=3.0 in]{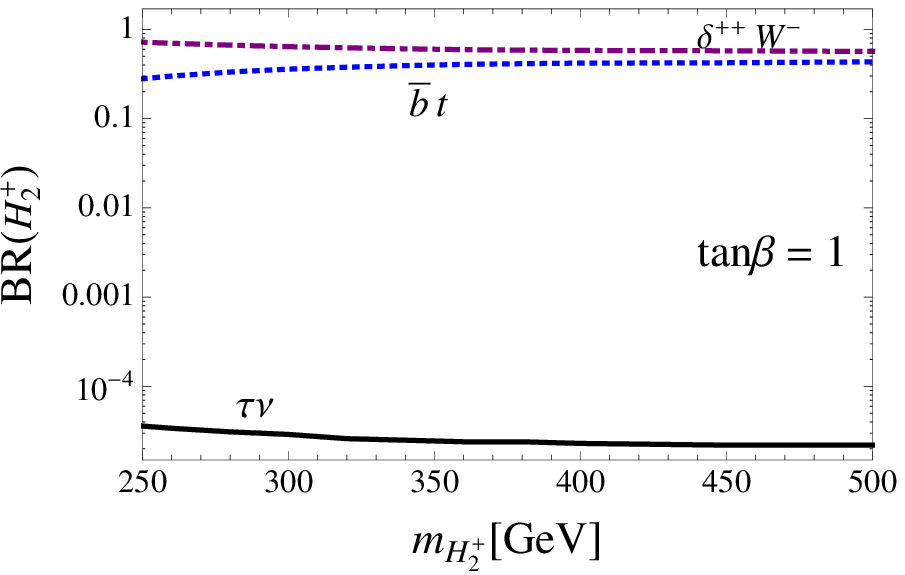} ~\includegraphics[width=3.0 in]{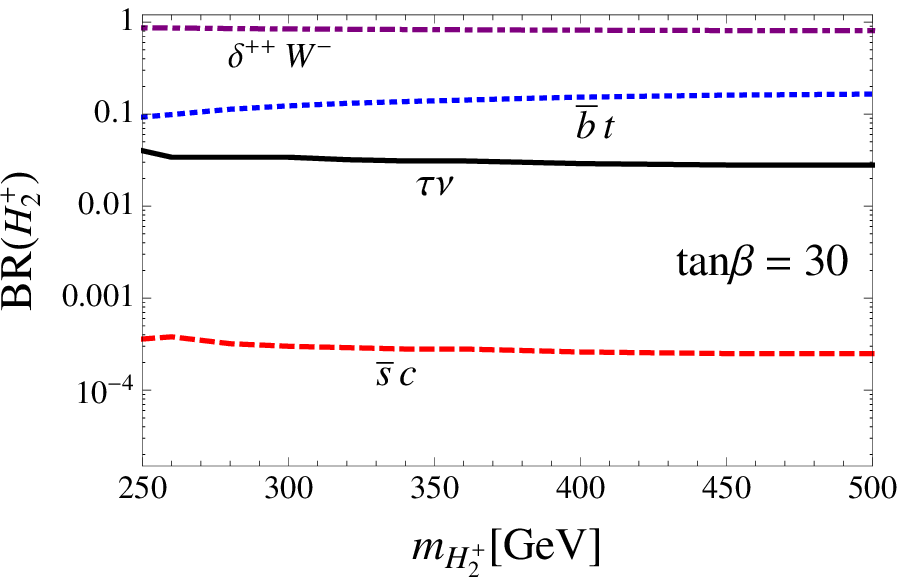}
\end{center}
\caption{ The legend is the same as that in Fig.~\ref{fig:BrHc} but for $H^+_2$. 
\label{fig:BrHc2}}
\end{figure}

Now we  realize that the doubly charged Higgs  dominantly decays into one charged Higgs and one $W$ gauge boson in our model. For simulating the $\delta^{\pm\pm}$ events, we further discuss the decays of $W$ and $H^\pm_{1,2}$. 
 Since the decays of $W$ boson  are clear in the SM, we just focus on the  $H^\pm_{1,2}$ decays. According to Eq.~(\ref{eq:hff}), we see that $H^\pm_{1,2}$ could decay to leptons and quarks, in which the couplings to fermions are proportional to the masses of fermions. By neglecting the small mass effects and CKM suppressions, we present the BRs for $H^+_1$ decays as a function of $m_{H^\pm_1}$ in Fig.~\ref{fig:BrHc} with $\tan\beta=1$ (left panel) and $\tan\beta=30$ (right panel). 
For heavier charged Higgs $H^\pm_2$, besides the decay channels appearing in $H^{\pm}_1$, the decay $H^\pm_2\to \delta^{\pm \pm} W^\mp$ can also occur with  our  parameter settings. Hence, the BRs for $\tan\beta=1$ (left panel) and $\tan\beta=30$ (right panel) as a function of  $m_{H^\pm_2}$ are given in Fig.~\ref{fig:BrHc2}. By the plot, we see clearly that $H^+_2\to (\delta^{++} W^{-}, t \bar b)$ are the main decay modes and the BR of former is larger than that of latter. Nevertheless,  it is worthy to mention that  the off-shell $H^\pm_2$ generated in $\delta^{\pm\pm}$ decays will convert into $t \bar b (\bar t b)$.
 
 \section{Simulation studies}

In this section, we discuss the possible signal/background events, the cuts for event selections and the significance for discovering the doubly charged Higgs. In order to generate the simulated events, we employ the event generator {\tt MADGRAPH/MADEVENT\,5}~\cite{Ref:MG}, where the necessary Feynman rules and relevant parameters of model are created by FeynRules 2.0 \cite{Alloul:2013bka}. We  use   {\tt PYTHIA\,6}~\cite{Ref:Pythia} to deal with the fragmentation of hadronic effects,  the  initial-state radiation (ISR) and final-state radiation (FSR) effects, and the decays of SM particles e.g. $W$-boson, $t$-quark, etc. 
In event generation, we use the {\tt NNPDF23LO1} PDFs~\cite{Deans:2013mha}.
In addition, the generated  events are also run though the {\tt PGS\,4} detector simulation~\cite{Ref:PGS}. In following analysis,  we take collision energy at 13 TeV and the integrated luminosity is 40 fb$^{-1}$, which could be reached after 1 year running at 13 TeV~\cite{CMS:2013xfa, ATLAS:2013hta}. The results for 14 TeV should be similar.

\subsection{Signals and backgrounds}

The unique character of $\delta^{\pm\pm}$ is carrying two electric charges. For searching for the signals of $\delta^{\pm\pm}$, we require that the generated events at each collision have  the same-sign  charged lepton pairs $\ell^\pm \ell^\pm$ ( $\ell = e, \mu$) in the final states.  Unlike the cases in CTTSM and Georgi-Machacek model \cite{Chiang:2012dk,Georgi:1985nv}, where the same-sign dileptons are produced 
 by $\delta^{\pm \pm}$ decays of leptonic channels directly or $W^\pm W^\pm$ channel, 
the production of $\ell^\pm \ell^\pm$ in our model is through more intermediate states. As mentioned before, the main decay channels of doubly charged Higgs are
\begin{align}
\delta^{\pm \pm} \rightarrow W^{\pm^*} H_{1}^\pm,  \ W^{\pm^{(*)}} H_{2}^{\pm^*}\,. \label{eq:lep1}
\end{align}
With the parameter settings in Eq.~(\ref{eq:setting}),  the condition for the on-shell or off-sell $W$-boson depends on the mass of $\delta^{\pm\pm}$. Thus, one of the two same-sign leptons is emitted from this $W$-boson, e.g. $W^{(*)} \to \ell \nu_\ell$. 

Furthermore, according to the interactions in Eq.~(\ref{eq:hff}) and Table~\ref{tab:interactions}, we find that up to three-body decays, the dominant decay modes of $H^\pm_{1}$ and $H^{\pm*}_2$  are  
 \be
&&H^{{+(-)[*]}}_{1[2]} \to t^* \bar b \; (\bar t^* b) \to b W^+ \bar b ( \bar b W^- b)\,. \label{eq:lep2}
 \ed
Therefore, the other lepton of the same-sign dilepton is from the on-shell $W$-boson which is emitted by top-quark.  Since the same-sign dilepton from $\delta^{\pm\pm}$  are dictated by the processes shown in Eqs.~(\ref{eq:lep1}) and (\ref{eq:lep2}), the  kinematical distributions of the two leptons should be different from other Higgs triplet models.  We will show the differences later.   As to other particles produced during $pp$ collision, we require them to convert into jets. Since there are more than four jets  in the final states, the searching signals for $\delta^{\pm\pm}$ are set to be
\begin{equation}
\ell^\pm \ell^\pm + {\rm four \, or \, more \, jets}\,.  \label{eq:sig}
\end{equation}
%
%
%

The background events from the SM  could  mimic the signals of Eq.~(\ref{eq:sig}). 
For analyzing the  backgrounds, we classify the  possible processes as Drell-Yen (DY), EW, QCD, top and $VV$ ( $V=W$ or $Z$) backgrounds and write them as follows: 
\begin{enumerate}
\item DY background : $p p \rightarrow l^+ l^-$(+ISR/FSR)
\item EW background : $p p \rightarrow W^\pm W^\pm j j (\alpha^4)$
\item QCD background : $p p \rightarrow W^\pm W^\pm j j (\alpha^2 \alpha_s^2)$
\item top background : $p p \rightarrow W^\pm t \bar{t}$, $p p \rightarrow W^\pm t \bar{t} j$   
\item VV(V= W or Z) background : 
$p p \rightarrow W^\pm Z +n j$,  $p p \rightarrow Z Z+ nj$, 
\end{enumerate}
where the number of jets $n$ for VV backgrounds is taken as $n \leq 2$~\cite{Chiang:2012dk}. $W^\pm W^\pm +nj$ events  in $VV$ background have been included in EW and QCD background, therefore they should be excluded. 
%
%
Although  DY processes  in principle could contribute to the background, since the second highest transverse momentum of the same-sign dilepton and transverse momenta of jets are small, their contributions indeed are negligible. 
%
%
%
We thus ignore the DY background in the simulation analysis.

\subsection{Kinematical cuts}

For enhancing the signals of $\delta^{\pm\pm}$ and reducing the possible backgrounds, we need to propose some strategies of kinematical cuts. For excluding the soft leptons and jets, when we generate the events by event generator, 
%
we set the preselection conditions for leptons and jets to be 
\begin{align}
\label{eq:cuts_basic}
& p_T (\ell) > 10 \ {\rm GeV}, \quad \eta(\ell) < 2.5, \nonumber  \\
& p_T (j) > 20 \ {\rm GeV}, \quad \eta(j) < 5.0, 
\end{align}
where $p_T$ is the transverse momentum and $\eta = 1/2 \ln (\tan \theta/2)$ is pseudo-rapidity  with $\theta$ being the scattering angle  in the laboratory frame.

\begin{figure}[hptb] 
\begin{center}
\includegraphics[width=4.0 in]{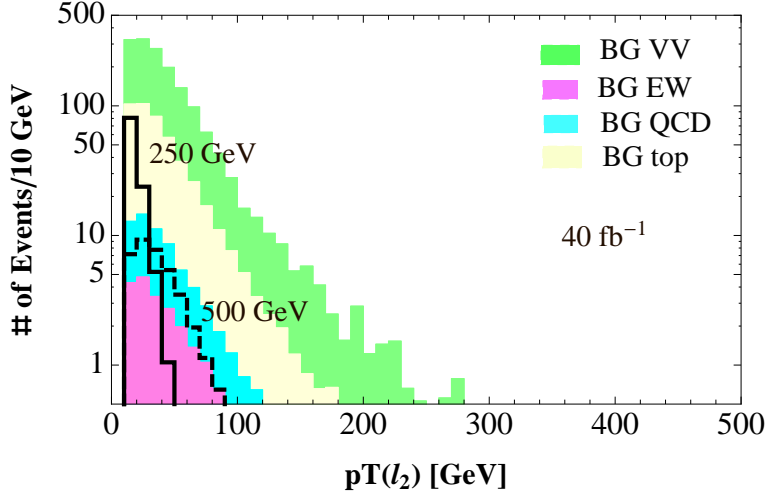} 
\end{center}
\caption{  Histograms of signal and background events versus  $p_T(\ell_2)$, where  the cuts of  Eq.~(\ref{eq:cuts_basic}) and  $m_{\delta^{\pm\pm}}=(250, 500)$ GeV are adopted.  For normalizing the histograms, we use the luminosity of 40 fb$^{-1}$. 
\label{fig:pTl2}}
\end{figure}

Since the signal processes have many b-jets in the final states, the number of b-tagging is a useful criterion to reject the backgrounds. In addition, differing from the CTTSM and Georgi-Machacek model that both same-sign charged leptons have larger $p_T$,  due to the small mass difference between $\delta^{\pm \pm}$ and $H_1^\pm$ in the parameter settings, the charged lepton from the decay of $\delta^{\pm \pm} \rightarrow \ell^\pm \nu H_{1}^\pm$ has a lower $p_T$. For understanding clearly, we plot the histograms of events versus the transverse momentum of the second highest $p_T$ lepton  in Fig.~\ref{fig:pTl2}, where we take  $m_{\delta^{\pm\pm}}=(250, 500)$ GeV and use the luminosity of 40 fb$^{-1}$.  It is clear that  the second highest $p_T$ leptons of signal events  prefer to locate at small $p_T$. 
%
Therefore, when we collect the events that are  run through Pythia and PGS detector simulation,  we further employ the new conditions for event selection as
\begin{equation}
\label{eq:cuts_addtional}
N_{\rm b-jet} \geq 1, \quad p_T(\ell_2) < 60 \ {\rm GeV}\,, 
\end{equation}
where $N_{\rm b-jet}$ denotes the number of b-jet, $\ell_2$ stands for the second highest $p_T$ charged lepton and the upper limit of $p_T(\ell_2)$ is referred to  the distributions in Fig.~\ref{fig:pTl2}.

\begin{figure}[hpbt] 
\includegraphics[width=3.0 in]{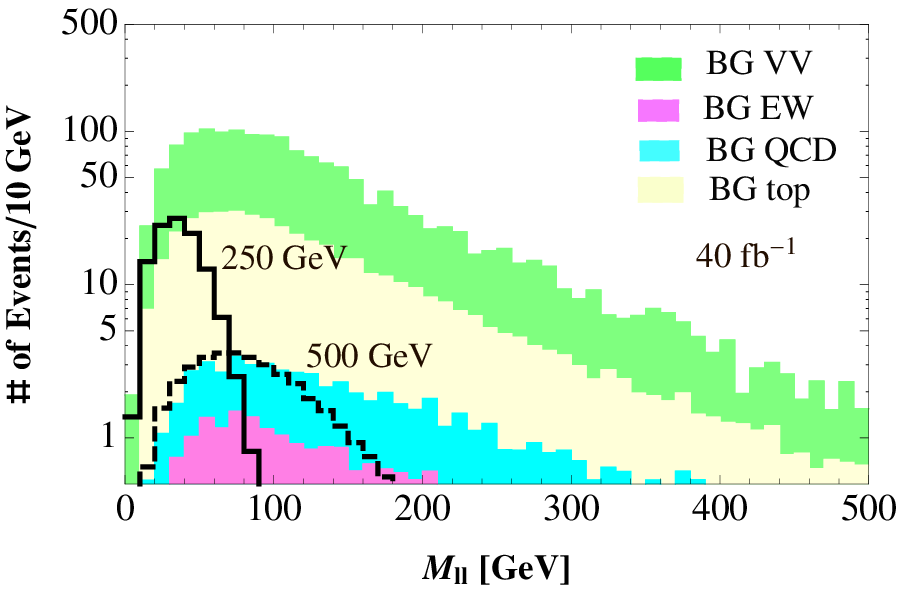} 
\includegraphics[width=3.0 in]{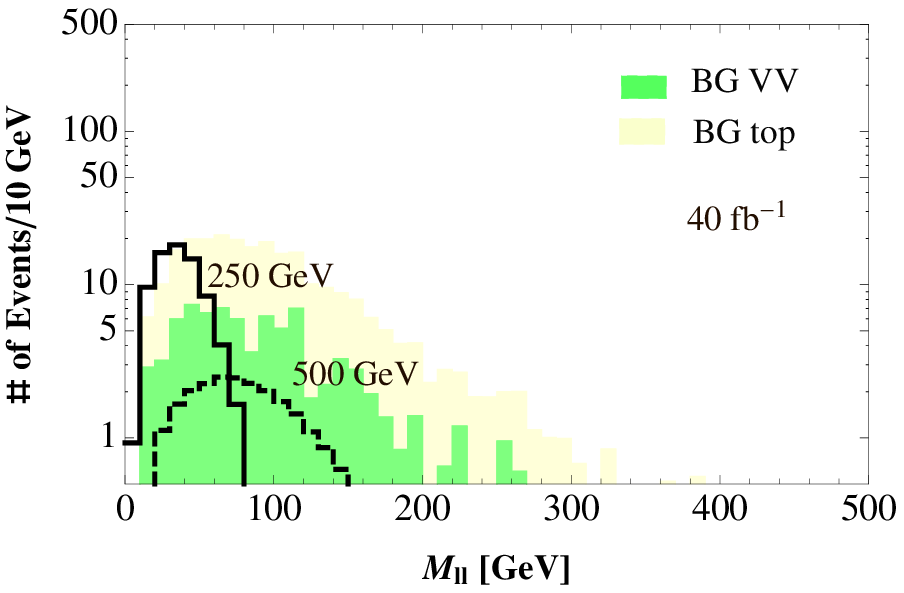} 
%
\caption{  Distributions of invariant mass of the same-sign dilepton with the basic cuts of Eq.~(\ref{eq:cuts_basic}) (left panel) and the additional cuts of Eq.~(\ref{eq:cuts_addtional}) (right panel). For illustration, we take $m_{\delta^{\pm\pm}} =(250, 500)$ GeV and the luminosity of 40 fb$^{-1}$. 
\label{fig:IMll}}
\end{figure}

Besides $N_{\rm b-jet}$ and $p_T(\ell_2)$, we also find that it is a useful method to reduce the backgrounds if we survey the invariant mass of the two same-sign leptons, denoted by $M_{\ell^\pm \ell^\pm}$. As discussed before, the same-sign leptons are generated through  multiple intermediate states in $\delta^{\pm\pm}$ decays. It is expected that the major values of $M_{\ell^\pm \ell^\pm}$ are not large. We present the distributions of dilepton invariant mass  for signals and backgrounds in Fig.~\ref{fig:IMll},  where the left panel results from the  cuts of Eq.~(\ref{eq:cuts_basic}) and the right panel is  arisen from the further cuts of Eq.~(\ref{eq:cuts_addtional}). By the plots, we see that  the signal events tend to locate at small region of  the invariant mass. Consequently, we adopt  the proper kinematical cut for $M_{\ell^\pm \ell^\pm}$  as
\begin{equation}
M_{\ell^\pm \ell^\pm} < \frac{m_{\delta^{\pm \pm}}}{4}. \label{eq:Mll}
\end{equation}
%
Since the invariant mass distribution of signal does not have a peak at the mass of doubly charged Higgs, for extracting the mass value of $\delta^{\pm\pm}$,  one needs to perform the fitting to the entire distribution with sufficient statistics.

\subsection{Discovery potential} 

After establishing the criteria for event selection, we start to calculate  the number of  signals and each background events and  investigate the  resulting significance. In our calculations, the significance is defined by~\cite{Ball:2007zza}  
\begin{equation}
S = \sqrt{2[(n_s+n_b)\ln (1+n_s/n_b)-n_s]},
\end{equation}
where $n_s$ and $n_b$ denote the number of signal and background events, respectively. For illustration,  we take $m_{\delta^{\pm \pm}}=200$ GeV and the integrated luminosity is set to be 40 fb$^{-1}$. Accordingly,  after employing the kinematical cuts, the number of various events  is shown in Table~\ref{tab:Nevents}. By the table, we see clearly that the condition with  $N_{\rm b-jet}\geq 1$ indeed can  significantly eliminate the backgrounds, especially in the $VV$ background. Furthermore, by using the cut of $M_{\ell^+ \ell^+}$ proposed in Eq.~(\ref{eq:Mll}), we find that the strongest competitor of signal is from the top background,  in which the produced final states are similar to those from $\delta^{\pm\pm}$ decays. 
%
%
%
\begin{table}[hptb]
\begin{ruledtabular}
\begin{tabular}{lcccccc}
cuts & signal & EW & QCD  & $t \bar{t}$ & VV & S \\ \hline \hline
Basic cuts & 81.2 & 22.3 & 44.2 & 398. & 1095. & 2.04 \\ \hline
b-tagging & 48.9 & 1.42 & 3.90 & 216. & 92.6 & 2.69  \\ \hline
$p_T(\ell^+_2) < 60$ GeV & 48.9 & 1.17  & 3.23 & 180. & 72.8 & 2.96 \\ \hline
$M_{\ell^+ \ell^+} < 50$ GeV & 46.4 & 0.13 & 0.48 & 33.6 & 17.7 & 5.71 
\end{tabular}
\caption{Number of signal and background events when the proposed  kinematical cuts are applied, where we have used the luminosity of 40 fb$^{-1}$, $m_{\delta^{++}} =200$ GeV and $\tan \beta =1$. Both "$\ell^+ \ell^+$+jets" and "$\ell^- \ell^-$+jets" events are  included. \label{tab:Nevents}}
\end{ruledtabular}
\end{table}

\begin{figure}[hbt] 
\includegraphics[scale=0.8]{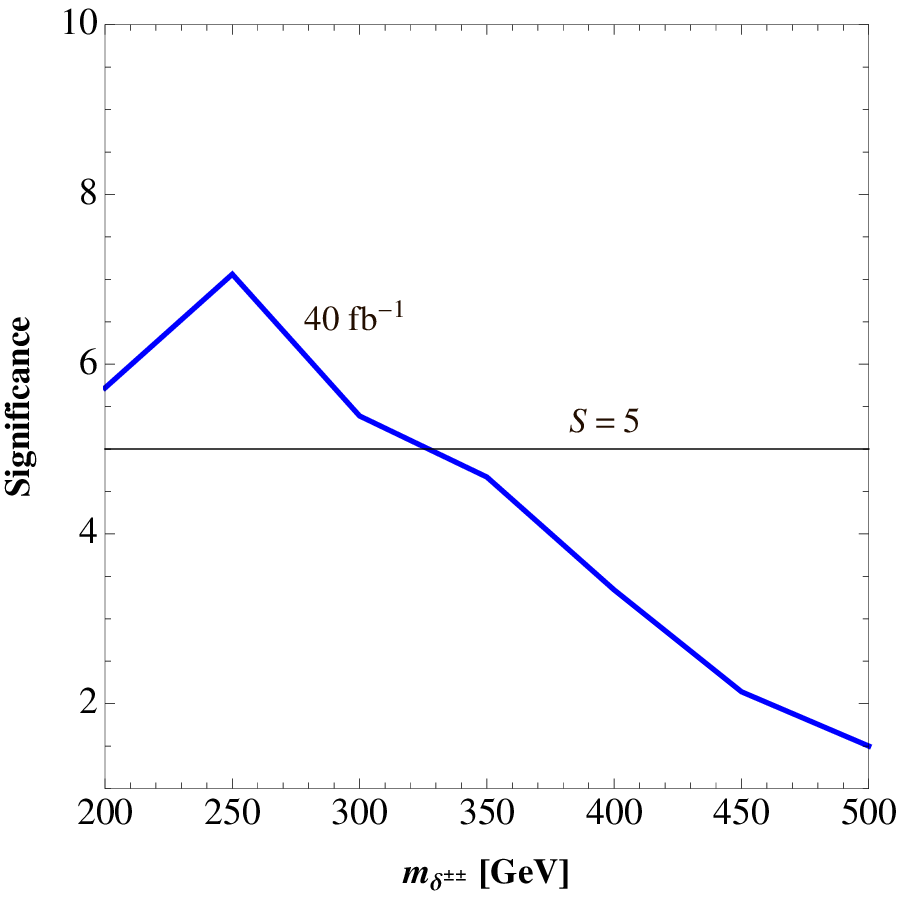} 
\includegraphics[scale=0.8]{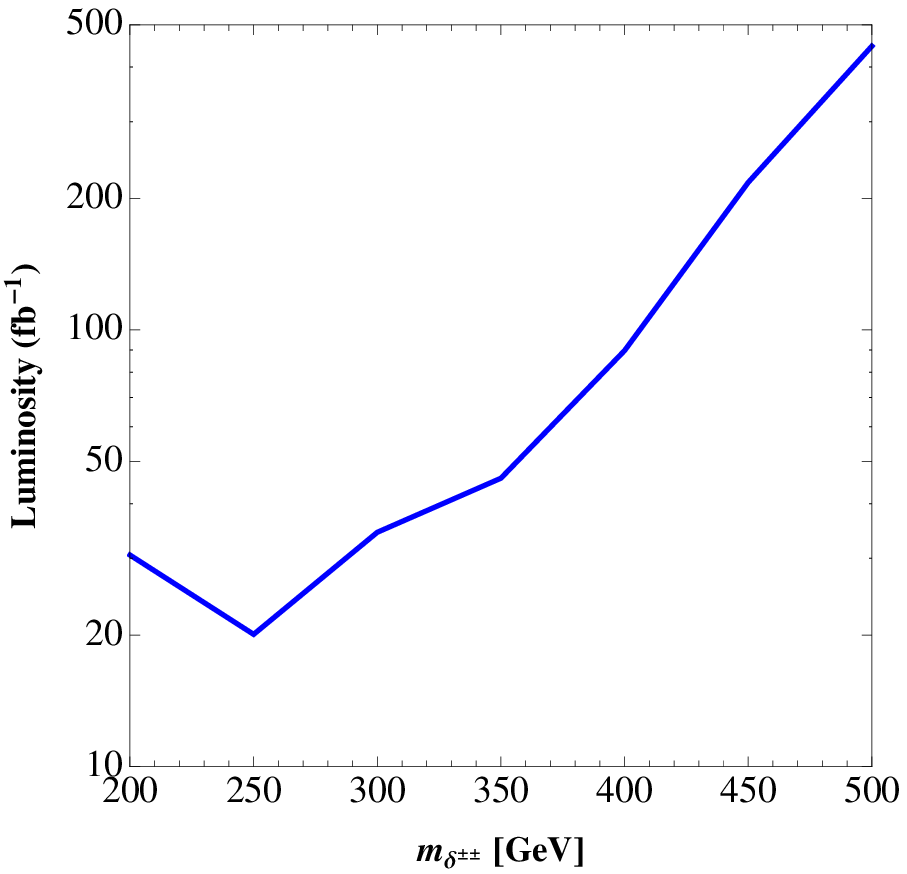} 
\caption{  Significance with 40 fb$^{-1}$ (left) and luminosity for $S > 5$ (right) as a function of $m_{\delta^{\pm\pm}}$.  The collision energy of 13 TeV is applied for both plots. 
\label{fig:significance}}
\end{figure}

In order to understand how the significance depends on the mass of $\delta^{\pm\pm}$ and what the value of luminosity is necessary to produce a 5$\sigma$ observation of doubly charged Higgs, we plot the related results in Fig.~\ref{fig:significance}.  
The left (right) panel  is the estimated significance (luminosity)  as  a function of $m_{\delta^{\pm\pm}}$. By the figure, 
one can find that  the doubly charged Higgs boson with a mass lower than 330 GeV can be discovered at the LHC with an integrated luminosity of 40 fb$^{-1}$. Additionally,  the doubly charged Higgs boson with a mass of 450 GeV can be discovered at the LHC with an integrated luminosity of 300 fb$^{-1}$  which is a target luminosity of LHC at 13-14 TeV energy by the end of 2021~\cite{CMS:2013xfa, ATLAS:2013hta}.
%

\section{ Summary}

We have studied the new properties of doubly charged Higgs and its discovery potential at the LHC in the THD extension of conventional type-II seesaw model.  We find that the new dimension-3 interactions $\mu_j H^T_j i\tau_2 \Delta^\dagger H_k$ appearing in the scalar potential  lead to a fantastic mixing effect between the singly charged Higgses of Higgs doublet and triplet. The mixing results completely different decay patterns in $\delta^{\pm\pm}$. 

With small leptonic Yukawa couplings, $Y_{\ell' \ell} \ll 1$, and $v_\Delta/v \ll 1$,  due to the mixing effects, the doubly charged Higgs mostly  decays into $W^{\pm^*} H^\pm_1$ and $W^{\pm^{(*)}} H^{\pm^*}_2$, but not  directly into $\ell \ell$ and $WW$ modes  which are usually discussed in the literature. That is, the search for doubly charged Higgs through either large $Y_{\ell' \ell}$ or large $v_\Delta$ in experiments should be reanalyszed by the new decay channels. According to our analysis,  it is found that in the considered model QCD processes  are the predominant effects to produce the $\delta^{\pm\pm}$, read as $pp \to H_2^+ \bar t b (H_2^- t \bar b) \to \delta^{++} W^- \bar t b (\delta^{--} W^+ t \bar b)$ and $\bar b(b) g \to H_2^+ \bar t (H_2^- t) \to \delta^{++}W^- \bar t (\delta^{--} W^+ t)$, while other Higgs triplet models  are arisen from EW processes.

For searching for the signals of $\delta^{\pm\pm}$, besides the preselection cuts imposed  in Eq.~(\ref{eq:cuts_basic}), in order to  further reduce the background events and enhance the significance of signal, we also propose the kinematical cuts on the number of b-jets, $p_T(\ell_2)$ and the invariant mass of same-sign dilepton, defined in Eqs.~(\ref{eq:cuts_addtional}) and (\ref{eq:Mll}). We find that with luminosity of 40 fb$^{-1}$ and collision energy of 13 TeV, $\delta^{\pm\pm}$ with mass below $330$ GeV could be observed at the $5\sigma$ level. Additionally, the observed mass of $\delta^{\pm\pm}$ could be up to 450 GeV when the luminosity approaches 300 fb$^{-1}$. \\

\noindent{\bf Acknowledgments}

 This work is supported by the Ministry of Science and Technology of 
R.O.C. under Grant \#: MOST-103-2112-M-006-004-MY3 (CHC) and MOST-103-2811-M-006-030 (TN). We also thank the National Center for Theoretical Sciences (NCTS) for supporting the useful facilities.

\end{document}